\documentclass{ws-procs9x6}

\begin{document}

\title{A high resolution scintillating fiber tracker with SiPM readout for 
the PEBS experiment}

\author{H. Gast, R. Greim, T. Kirn, G. Roper Yearwood$^*$, S. Schael}

\address{I. Physikalisches Institut B, Rheinisch-Westf\"alische Technische
 Hochschule Aachen,\\
Aachen, 52062, Germany\\
$^*$E-mail: roper@physik.rwth-aachen.de\\
www.physik.rwth-aachen.de}

\begin{abstract}
Using thin scintillating fibers with Silicon Photomultiplier (SiPM) readout a modular high-resolution charged-particle tracking detector has been designed. The fiber modules consist of 2 x 5 layers of 128 round multiclad scintillating fibers of 0.250mm diameter. The fibers are read out by four SiPM arrays (8mm x 1mm) each on either end of the module.

The basic features of this detector concept have been evaluated in a test beam in October 2006 using novel SiPM detectors with improved photon detection efficiency. This detector has been developed for a balloon borne spectrometer (PEBS) to measure the comsic ray positron- and electron flux with high precision.

This particle detection concept is also very interesting for future applications, for example as an outer layer of an ILC detector or for other astroparticle physics experiments.
\end{abstract}


\bodymatter

\section{Introduction}\label{aba:sec1}
Scintillating fiber trackers have already been realized using 
\textit{Visible-Light Photon Counters}\cite{dzero} or \textit{Multi-Anode 
Photomultiplier Tubes}\cite{nahnhauer} as photodectors. Multi-anode PMTs
do not exhibit a sufficiently high photodetection efficiency (generally around 
$20\%$ to $30\%$) to encourage their use for a high precision SciFi tracker 
using very thin scintillating fibers which yield only a small number of photons
for a traversing minimal ionizing particle. VLPCs on the other hand require a 
significant overhead since they require operating temperatures of $7K$. 

\begin{figure}
\begin{center}
\begin{tabular}{cc}
\epsfig{file=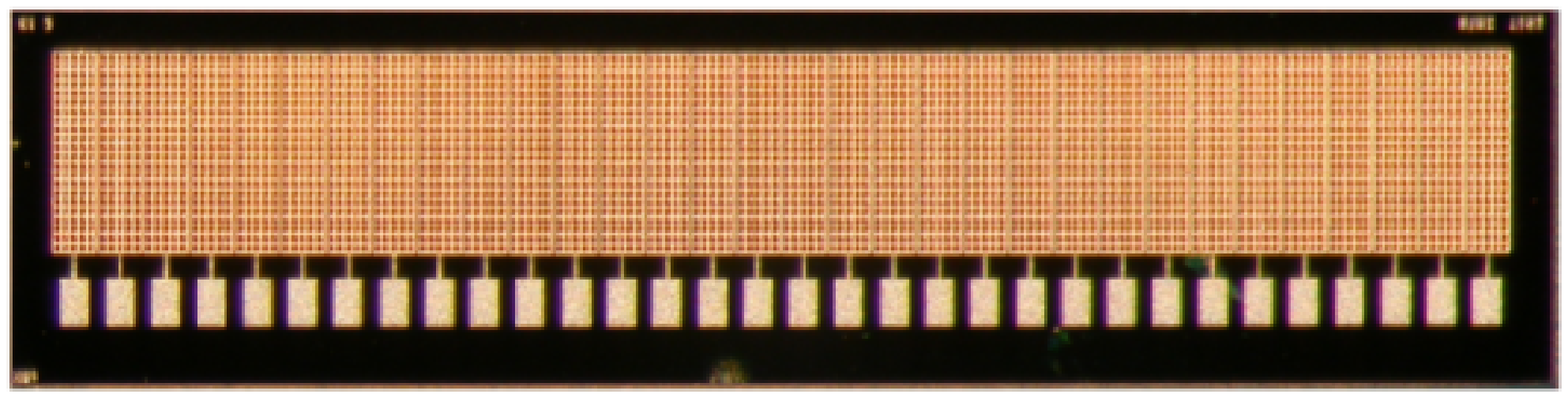, angle=270, width=0.6\textwidth}
&
\epsfig{file=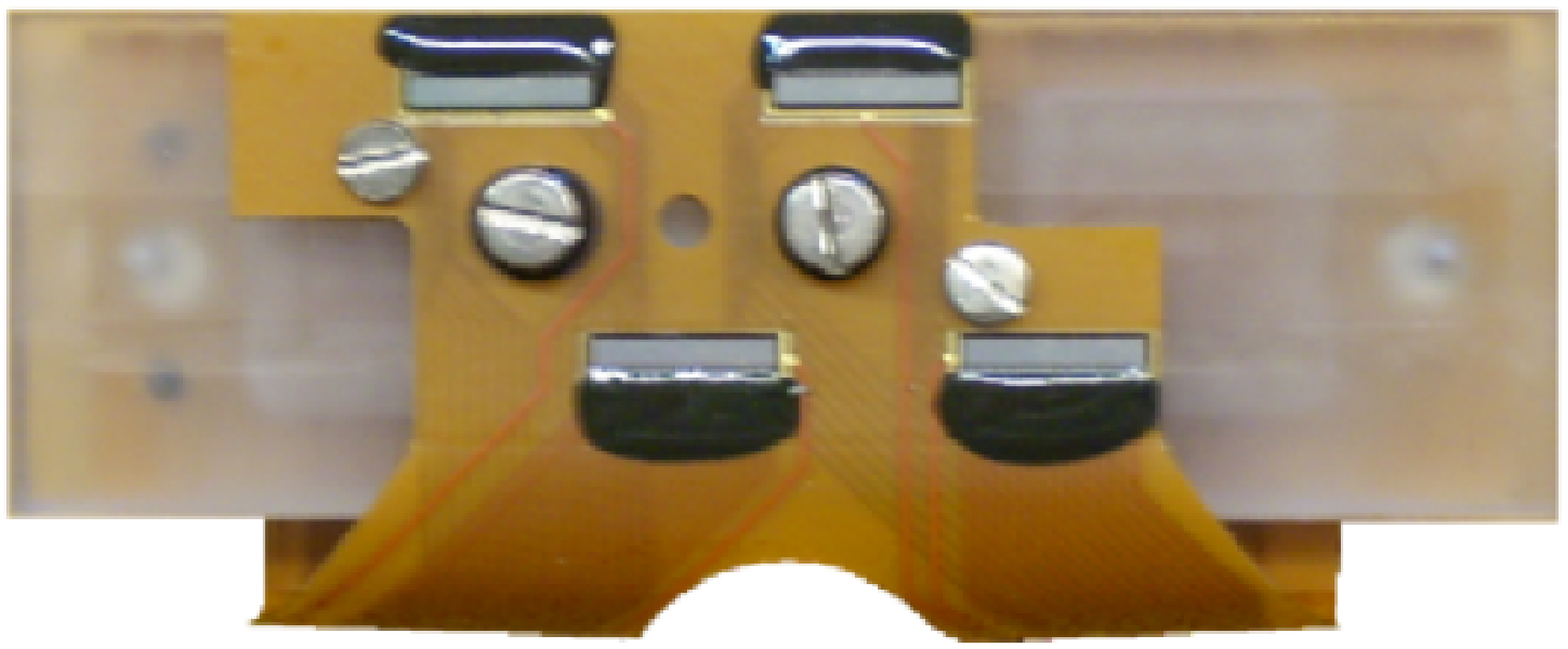, angle=270, width=0.3\textwidth}
\end{tabular}
\caption{Photography of a working SiPM array prototype from INFN Perugia (left) and 4 SiPM already bonded and ready to read out a tracker module (right)}
\label{figsipmarray}
\end{center}
\end{figure}

SiPMs achieve high photodetection efficiencies (PDE) of over $60\%$\cite{mppc} 
and allow for operation at room temperature ($> 20^{\circ}C$). Furthermore 
novel SiPM arrays with very densely packed channels allow for 
a compact design (see fig. \ref{figsipmarray}). 

\section{SiPM}
A SiPM is a multipixel semiconductor photodiode that achieves a high intrinsic gain of $10^5$ to $10^6$ when being operated above its breakdown voltage\cite{dol01}. In addition, they scale to small dimensions, allowing for a compact readout of thin scintillating fibers. 

Commercial distributors of SiPMs are Hamamatsu\cite{ham07}, Japan and Photonique\cite{photoniqueweb}, Switzerland. Photonique SiPMs of type SSPM-0606EXP ($40\%$ PDE) and SSPM-050701GR ($25\%$ PDE) were used during the beamtest of a SciFi/SiPM tracker prototype (fig. \ref{figpde}). New SiPMs of type Hamamatsu MPPC S10362-100C ($65\%$ PDE) and 32 channel SiPM arrays produced by the INFN Perugia are already available and currently tested. 

\section{Beamtest of the first Prototype}
The prototype consisted of $300\mu m \times 300 \mu m$ square, multiclad fibers of type Bicron\cite{bicron} BCF-20 with white EMA coating and Photonique SiPMs of type SSPM-050701GR and of type SSPM-0606EXP. The peak emission wavelength of BCF-20 fibers is at $492nm$, matching the peak sensitivity of Photonique SSPM-0606EXP SiPMs.

The scintillating fibers were arranged in two ribbons of $3 \times 10$ fibers. The fiber ribbons were stabilized using glue as an adhesive. Both ends of the 3-fiber ribbons were glued into a plastic connector and polished. One end was connected to a SiPM by mounting it into a copper block and held in place by an aluminum frame and a spring.
The SiPMs were mounted into the copper-block to allow for a temperature control. During part of the beam test  the opposing end of the fibers was covered by a reflective foil to increase the light output for the SiPMs.

A beam telescope with four silicon strip modules from the CMS tracker project was used to measure the position of the incident particles. 

The beamtest of the prototype took place in a 10GeV proton beam at PS, CERN. During the beamtest, 1.5 million events were recorded and about 800,000 particle tracks were reconstructed with the beam telescope. The position of each fiber column was determined by reconstructing the position of particles that produced a high signal within the fiber. The average measured distance between two fiber columns was $309\mu m$ with a precision of $10\mu m$. The spatial resolution for particles of perpendicular incidence that pass through all three fibers of one fiber column was about $90\mu m$ which matches the expected intrinsic resolution of $\frac{d}{\sqrt{12}}$ where $d$ is the fiber pitch.

Knowing the positions of the fibers, we determined the average photoelectron yield for particles that passed through one of the fiber columns. For particles with perpendicular incidence the average photoelectron yield for both types of SiPMs with and without reflective foil on the opposing fiber end was measured. The SSPM-050701GR signal was $1.9 \pm 0.3$ photoelectrons without and $3.2 \pm 0.2$ photoelectrons with reflective foil. The SSPM-0606EXP achieved an average photoelectron yield per MIP of $3.6 \pm 0.2$ without reflective foil and $5.4 \pm 0.3$ with reflective foil (omitting one SiPM that actually showed a reduced photoelectron yield after adding the reflective foil).

\begin{figure}
\begin{center}
\begin{tabular}{cc}
\epsfig{file=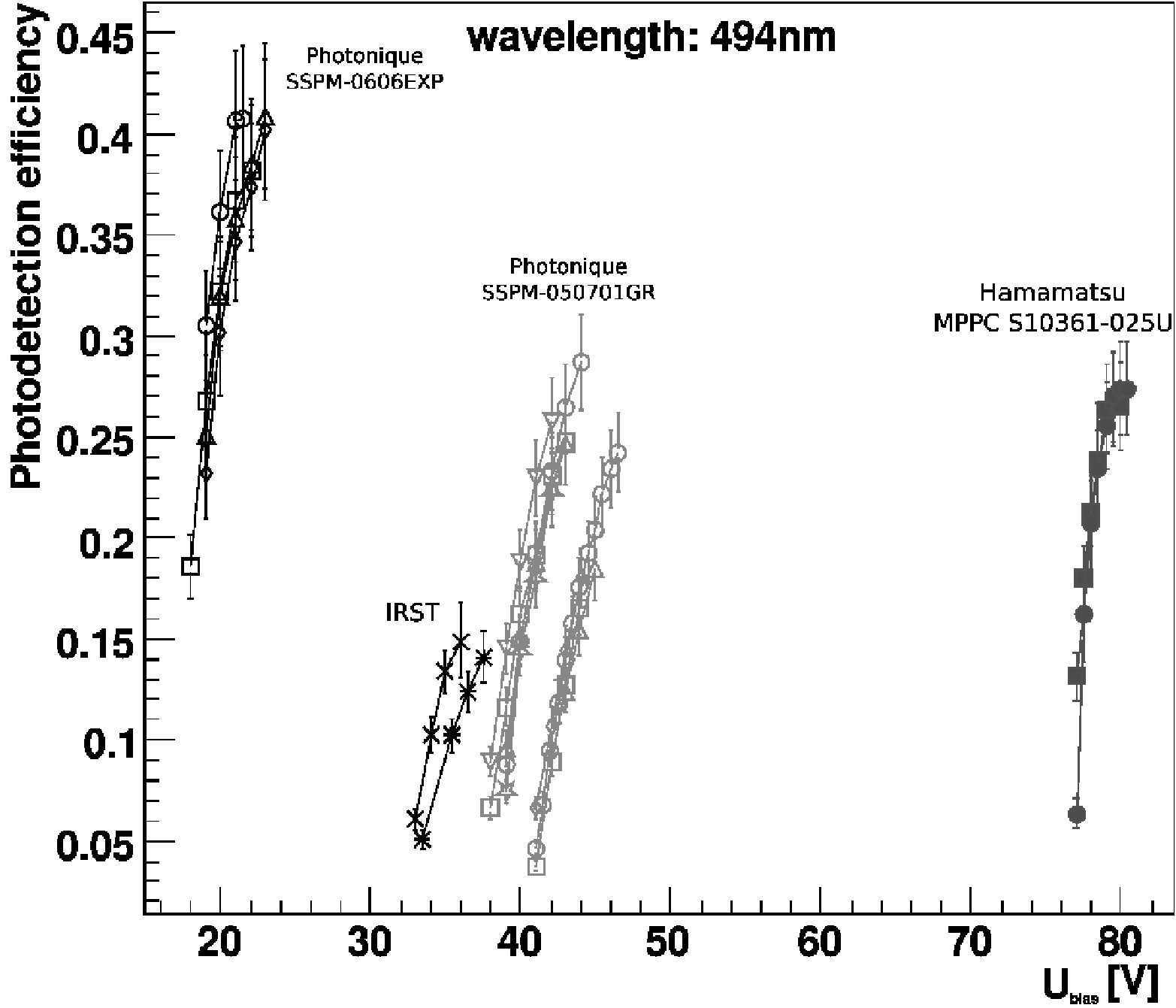, width=0.45\textwidth} & \epsfig{file=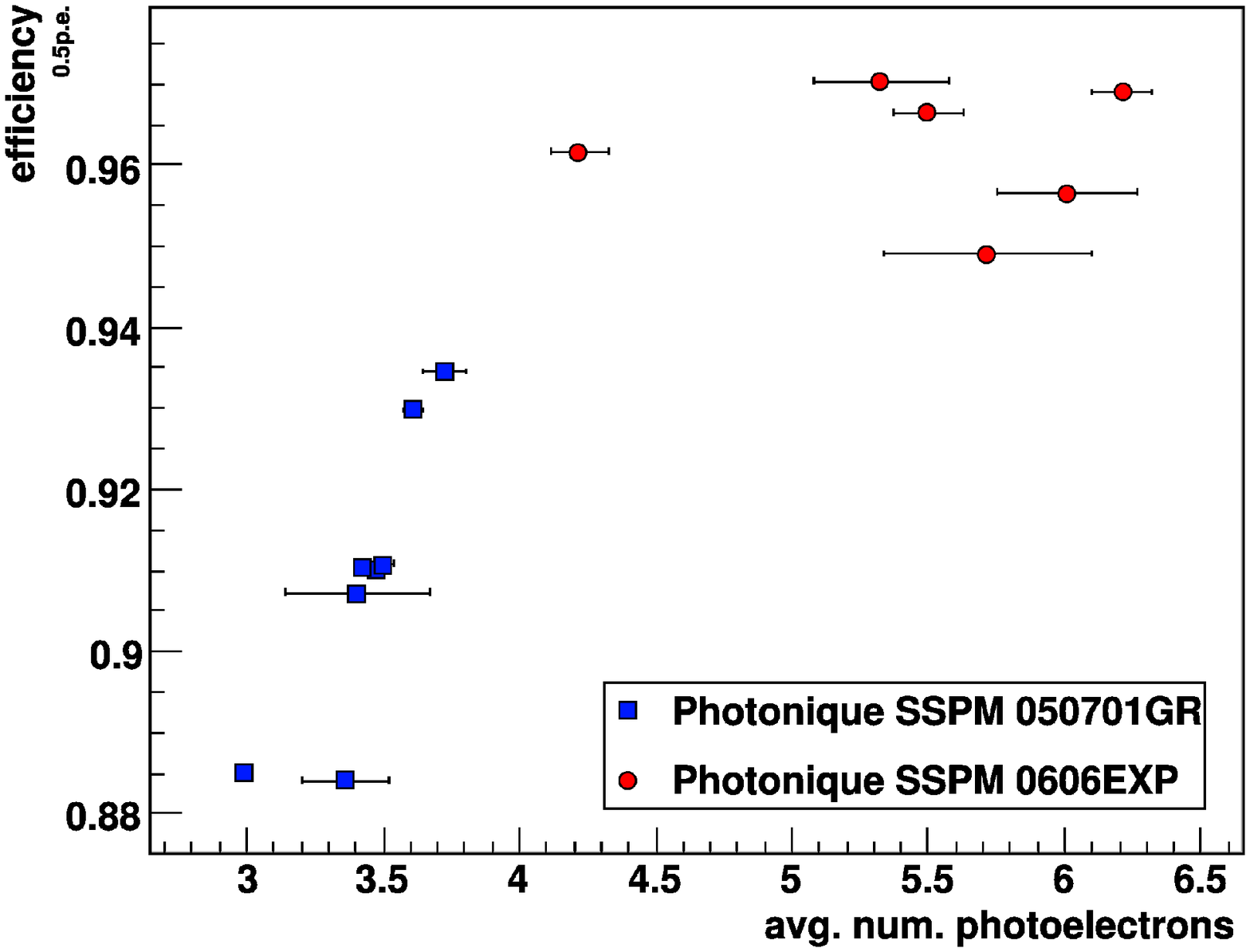, width=0.5\textwidth}
\end{tabular}
\caption{Measured photodetection efficiencies for various SiPM at 494nm (left). To the right plot of the efficiency of the particle detection (for the lowest possible cut at $0.5$ photoelectrons) vs. the average number of photoelectrons as they were measured during the beamtest.}
\label{figpde}
\end{center}
\end{figure}

The mean efficiency for perpendicular incidence, setting a cut at 0.5 photoelectrons, was $96\%$ for the ribbon read out by the SSPM 0606EXP and $91\%$ for the SSPM 050701GR (fig. \ref{figpde}). 

\section{Tracker Design}

\begin{figure}
\begin{center}
\begin{tabular}{cc}
\epsfig{file=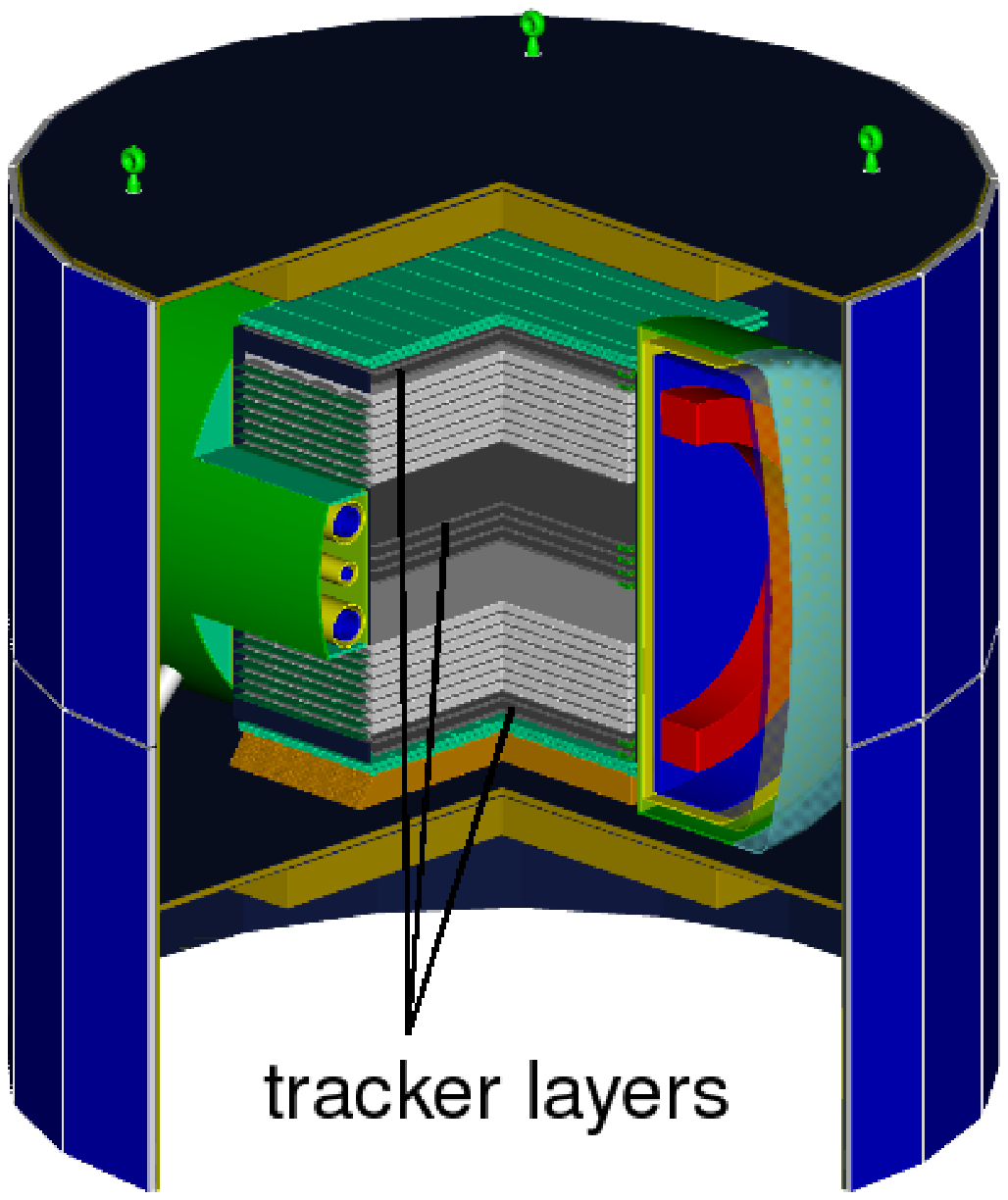, width=0.45\textwidth} &
\epsfig{file=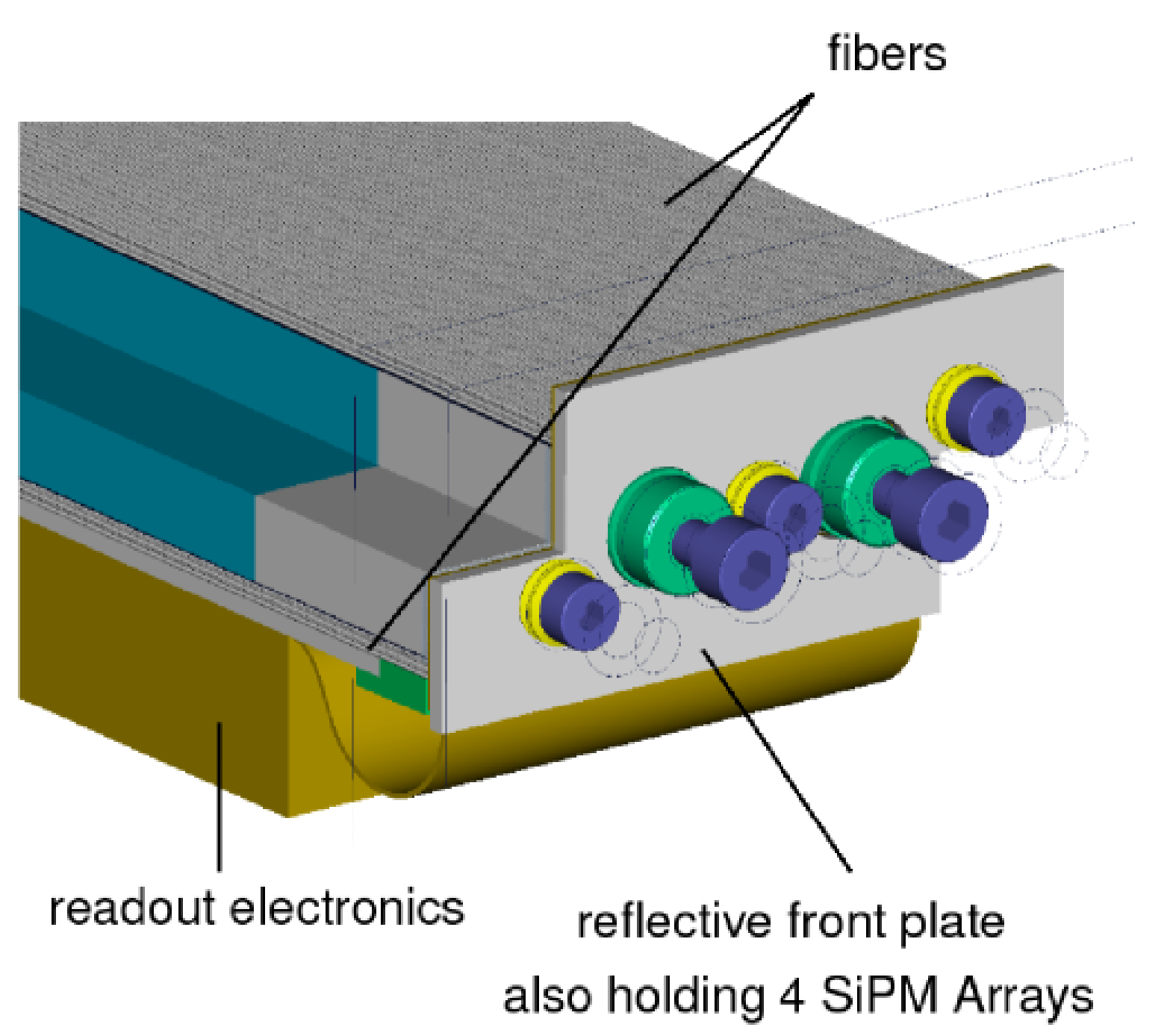, width=0.45\textwidth}
\end{tabular}
\caption{A schematic view of the complete PEBS detector (left) and a 3d view of a single module (right).}
\label{figpebs}
\end{center}
\end{figure}

The tracker design for the PEBS detector (fig. \ref{figpebs}) is modular. It consists of several layers of tracker modules, each module consisting of 10 layers of $0.25mm$ thin, round scintillating fibers with 128 fibers in every layer. 5 layers of fibers are glued to each side of a $10mm$ thick module core of Rohacell foam covered by $100\mu m$ thin carbon fiber skins on either side of the foam. Neighboring layers are shifted by one half of the fiber pitch with respect to each other to improve the spatial resolution. 

SiPM arrays with an area of $8mm \times 1.1mm$ and 32 readout cells, each sensitive cell covering an area of $0.23mm \times 1.1mm$, are used for column-wise fiber readout. The SiPM arrays are mounted on alternating ends of the fiber modules along with an integrated preamplifier and digitization solution. The respective opposing ends of the fibers are covered by a reflective coating. 

\begin{figure}
\begin{center}
\epsfig{file=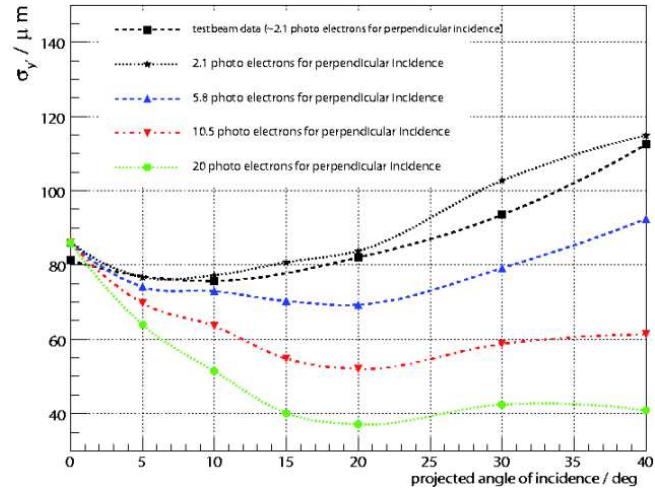, width=0.75\textwidth}
\caption{Spatial resolution for a bundle of fibers of $300\,\mu{}m$ width from testbeam data and Monte Carlo simulations. Testbeam data obtained with a fiber bundle without reflective foil and Photonique SSPM-050701GR SiPMs readout are plotted using square markers. Results from Monte Carlo simulations are added to study the behavior for improved photo electron yields.}
\label{figspatial}
\end{center}
\end{figure}

A dedicated Monte Carlo simulation, using the GEANT4\cite{geant4} package, has been developed for comparison to and generalization of the testbeam results\cite{gast}. A key question to be answered was the spatial resolution obtained with a fiber module as a function of the mean photoelectron yield $n_{p.e.}$ of the fiber-SiPM chain (see fig. \ref{figspatial}). For the photo electron yield achieved in the testbeam, a spatial resolution of $72\,\mu{}m$ is obtained at the mean projected angle of incidence, which is $\bar{\alpha}=11^\circ$ for the PEBS geometry.

\section{Conclusion}
The testbeam results indicate that this concept for a high-resolution SciFi/SiPM tracker is technically feasible. The average yield of 5.4 photoelectrons with a reflective foil on one fiber end and the SSPM-0606EXP SiPMs is acceptable. SiPMs with a reduced pixel density and a $50\%$ higher PDE are commercially available from Hamamatsu and will be tested thoroughly during the next test beam in fall 2007.
The light output from scintillating fibers can be improved by $20\%-40\%$ using fibers without white coating as measurements with different fiber coatings conducted for the CREAM experiment have shown\cite{cre05}. 

A spatial resolution as good as $40 \mu m$ is possible, depending on the granularity of the readout, the quality of the SiPMs, the qualtity of the optical coupling of the fibers to the SiPMs and the type of fibers used. For the final tracker an average spatial resolution of $60 \mu m$ is expected.

\end{document}